\documentstyle[aps,prl,multicol,epsf]{revtex}

\begin{document}
\title{Critical exponents at the ferromagnetic transition in
tetrakis(diethylamino)ethylene-C$_{60}$ (TDAE-C$_{60}$)}
\author{Ale\v{s} Omerzu$^{1,2}$, Madoka Tokumoto$^{2}$, Bosiljka
Tadi\'{c}$^{1}$ and
Dragan Mihailovic$^{1}$}
\address{$^{1}$Jozef Stefan Institute, Jamova 39, 1000\ Ljubljana,
Slovenia\\
$^{2}$Nanotechnology Research Institute, National Insitute of Advanced
Industrial Science and Technology (AIST), 1-1-1 Umezono, Tsukuba,
Ibaraki 305-8568, Japan}
\maketitle

\begin{abstract}
Critical exponents at the ferromagnetic transition were measured for the
first time in an organic ferromagnetic material
tetrakis(dimethylamino)ethylene fullerene[60] (TDAE-C$_{60}$). From a
complete magnetization-temperature-field data set near $T_{c}=16.1\pm 0.05,$
we determine the susceptibility and magnetization critical exponents $\gamma
=1.22\pm 0.02$ and $\beta =0.75\pm 0.03$ respectively, and the field vs.
magnetization exponent at $T_{c}$ of $\delta =2.28\pm 0.14$. Hyperscaling is
found to be violated by $\Omega \equiv d^{\prime }-d\approx -1/4,$
suggesting that the onset of ferromagnetism can be related to percolation of
a particular contact configuration of ${{\rm C_{60}}}$ molecular
orientations.
\end{abstract}

\pacs{PACS numbers:}

\begin{multicols}{2}
Molecular ferromagnetism, particularly when only electrons in $p$ orbitals
are involved in the magnetic interactions, is a relatively newly discovered
phenomenon. Of the few compounds discovered so far which display signatures
of proper ferromagnetic (FM) behavior\cite{Allemand,AFCO}, the most studied
has been TDAE-C$_{60}$, which - by virtue of its relatively simple synthesis
(at least in powder form) and high Curie temperature of T$_{c}=$16 K - has
been investigated by many groups. However, the magnetic properties of this
material are not straightforward, and measurements on powder samples have
lead to apparently conflicting proposals regarding its low-temperature state
ranging from superparamagnetism\cite{super} to spin-glass\cite{sg} as well
as ferromagnetism. More recently, low-field electron paramagnetic resonance
\cite{Blinc} and susceptibility measurements on high-quality single crystals
showed more conclusively that the material - if properly chemically and
thermally prepared \cite{nature} - display clear signs of a transition to a
ferromagnetic state at around 16 K, in agreement with the original
suggestion of its discoverers \cite{Allemand}.

The origin of the ferromagnetic exchange interaction between C$_{60}$
molecules in TDAE-C$_{60}$ has been studied theoretically by a number of
groups\cite{nature,Jap,Mizoguchi}. Recently it was discovered \cite{nature}
that in the FM $\alpha $-phase of TDAE-C$_{60}$ two different orientations
of C$_{60}$-molecules may occur at low temperatures (labeled $I$ and $II$).
These orientations can lead to different contact C$_{60}$ configurations
along the direction of closest contact ($c$-axis)\cite{nature}, which
profoundly affect the exchange interaction along $c$-axis\cite{Jap}. In the
non-magnetic $\alpha ^{\prime }$ phase, the 6--6 double bond (nearly) faces
the center of the hexagon on the neighboring molecule, whereas in the FM
phase, a number of additional different mutual orientations are possible.
But, among these, the alternating $I$--$II$ contact configuration - in which
the double bond on one molecule approximately faces the center of the
pentagon of its neighbor - was shown to be dominant in the ferromagnetic
state.

To confirm the observation of a proper FM state and investigate the
associated critical behavior, we report here the first measurements of the
critical exponents associated with the ferromagnetic transition of TDAE-C$%
_{60}$. The measurements, which are also the first for any organic system,
are found to give remarkably self-consistent values of the critical
exponents in agreement with behavior expected for a ferromagnetic transition
in a system with a certain degree of disorder. They confirm the essentially
FM behavior, and also give important insight into the interactions
responsible for the ferromagnetism in these unusual materials.

Single crystals of TDAE-C$_{60}$ where grown by a diffusion method as
described in \cite{growth}. For magnetic measurements, a number of crystals
from different growth batches were sealed into quartz tubes under helium.
Since the ``as grown'' crystals of TDAE-C$_{60}$ are in their $\alpha
^{\prime }$ modification which shows no low-temperature ferromagnetic
transition\cite{PRB}, they were annealed in order to transform them into the
ferromagnetic $\alpha $ modification. The annealing was carefully done
through several intervals of 1 hour at 70$^{\circ }$C, each of them followed
by a measurement of the low-temperature magnetic properties. The annealing
procedure was stopped at the point were the low-temperature saturation
magnetization of the crystals reached it's maximum, i.e., when the whole
sample had transformed into the $\alpha $ modification. Magnetic
measurements were performed with a Quantum Design MPMS SQUID magnetometer
which enables a temperature stability better than 10\ mK and a measurement
of magnetization with a relative error of less than 0.1\% and since the
magnetic response for crystals smaller than 3 mm across is point-like, the
problems associated with demagnetization factors is avoided. Furthermore,
the magnetization was found to be independent of orientation.

In order to determine accurately the critical temperature of the
ferromagnetic transition, $T_{c}$, and the critical exponent $\gamma $ which
defines the temperature dependence of the zero-field magnetic susceptibility
$\chi $ in the critical region just above the transition $\chi (T)\sim $ $%
\left( T/T_{c}-1\right) ^{-\gamma }$, we have measured the static magnetic
susceptibility in a temperature interval 16 - 17.6 K. $\chi $ was determined
as a slope of the magnetization versus field curve $M(H)$, through ten
equidistant points between -5 and 5 Oe every 50 mK. To determine $T_{c}$ and
$\gamma $ from the experimental results we plot the inverse logarithmic
derivative $(d\ln \chi /dT)^{-1}\sim -(T/T_{c}-1)/\gamma $ versus reduced
temperature $\epsilon \equiv (T-T_{c})/T_{c}$. By varying $T_{c},$ the data
are made to appear on a straight line pointing to the origin. The best fit
of the data is shown in Fig.\ 1a for $T_{c}=16.05$. From the slope we
determine the exponent as $\gamma =1.22\pm 0.02$, the error reflecting
variations in $T_{c}$ within the range $16.05$--$16.15$ where the accuracy
of fit can not be further improved.

\narrowtext
\begin{figure}[h]
\epsfxsize=78mm\epsffile[24 68 506 639]{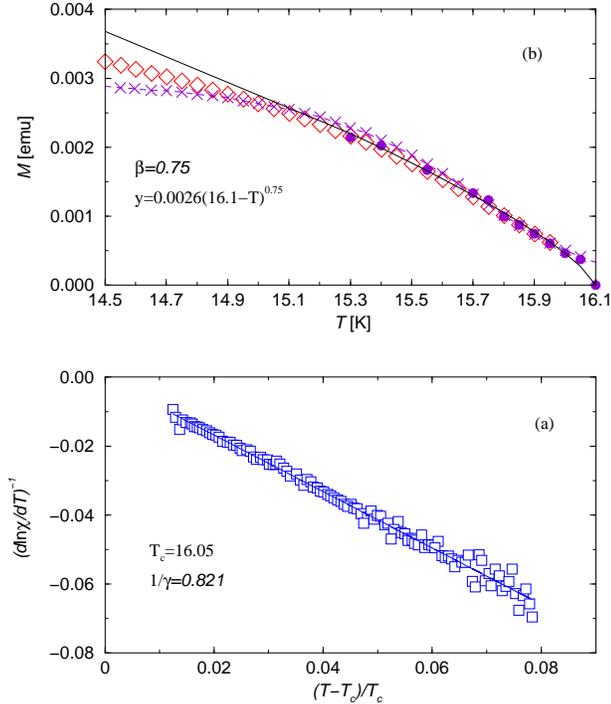}
\caption{(a)Inverse logarithmic derivative of the static susceptibility of
TDAE-C$_{60}$, $\left( d(\ln \protect\chi )/dT\right) ^{-1}\sim
-(T/T_{c}-1)/\gamma $ plotted against reduced temperature $%
(T-T_{c})/T_{c}$ in the critical region above $T_{c}$. (b) Temperature
dependence of the magnetization of TDAE-C$_{60}$ measured in a static
magnetic field of 20 Oe (open symbols) for temperatures below $T_{c}$. Also
shown are values of the magnetization measured at a field of 1 Oe (crosses)
scaled by a factor of 4.75 and extrapolated values for zero field (bullets)
scaled by factor of 6 along vertical axis. Solid line: the curve fit to $y=%
{\cal {B}}(T_{c}-T)^{{\protect\beta}} $, as indicated. }
\end{figure}
The critical exponent $\beta $ describes the temperature dependence of the
spontaneous magnetization, $M_{S}$, in the critical region $T\lesssim T_{c}$
through the relation $M_{S}\sim {\cal {B}}(1-T/T_{c})^{\beta }$. To
determine $\beta $ we have measured the magnetization at temperatures
between 14.4 K and 16 K in a low magnetic field. The data are shown in Fig.\
1b for 20 Oe. In fact, the {\em functional form} of the temperature
dependence of the magnetization at low fields and the temperature range
15.6--16.1 K does not depend on the value of the field. In this temperature
range and for small $H,$ different isotherms appear to be straight lines on a
$\log H$--$\log M$ plot (see Fig.\ 2). This implies that different $M(T)$
curves can be scaled to a unique functional form as Fig.\ 1b shows. The
scaled values of magnetization vs. temperature for lowest measured field $H$%
=1 Oe are seen to coincide with the ones measured at 20 Oe near $T_{c}$. In
addition, the extrapolations of the power-law isotherms in Fig.\ 2 to the $%
H=0 $ axis leads to the same $T$-dependence of the spontaneous magnetization
near the critical temperature, as is suggested by the susceptibility
measurements. (The corresponding curve scaled by a factor of 6 is also shown
in Fig.\ 2b.) Fitting the magnetization curve to the expected power-law
behavior, we find the best fit with $T_{c}=16.1\ $of $\beta =0.75$ within
statistical error bars of 0.03 (cf. Fig.\ 1b).

At $T=T_{c}$ the field dependence of magnetization follows the critical
isotherm $H\sim M^{\delta }$, where $\delta $ is the exponent for the
critical isotherm. To determine the critical exponent $\delta $ we have
measured the magnetization over four decades of magnetic fields between 1 Oe
and 10 kOe at several temperatures above and below $T_{c}$, as shown in
Fig.\ 2. By fitting several isotherms near the $T_{c}$ values suggested
above, we find the exponent $\delta $ in the range $\delta =2.14-2.41$, the
uncertainty reflecting the chosen value of $T_{c}$, giving $\delta =2.28\pm
0.14$.

An important feature of the measurements is that the critical exponents $%
\gamma $, $\beta $, and $\delta $ do not obey the scaling relation $\gamma
=\beta (\delta -1)$, which is expected to apply at a second-order phase
transition for a non-disordered system in equilibrium. Before we will
discuss the possible origin of the scaling violation, we show - as an
independent test of the consistency of the values for the critical exponents
- that the measured data obey a general scaling form \cite{Stanley,Sava}
\begin{equation}
M(T,H)\sim H^{1/\delta }M\left( \epsilon /H^{1/\beta \delta },1\right)
\label{scal_form}
\end{equation}
in the critical region at low fields $H\rightarrow 0$ and small relative
temperatures $\epsilon \equiv (T-T_{c})/T_{c}\rightarrow 0$. Eq.\ (\ref
{scal_form}) follows directly from the statement that the singular part of a
thermodynamic potential (or, equivalently, its derivatives) is a generalized
homogeneous function of its arguments. That is, $M(b^{\lambda _{T}}\epsilon
,b^{\lambda _{H}}H)=b^{\lambda }M(\epsilon ,H)$, by taking $b^{\lambda
_{H}}H\sim 1$ and the standard identification of the scaling exponents in
terms of $\beta $ and $\delta $ (see \cite{Stanley,Sava}). In Fig.\ 3 we
plot $M/H^{1/\delta }$ vs. $x\equiv \epsilon /H^{1/\beta \delta }$ using the
values of the critical exponents determined above and $T_{c}=16.1$. The
consistency in the exponents is demonstrated by a ``parallel'' collapse of
the curves for different $H$ values. As discussed in Ref.\ \cite{Sava} the
characteristic scaling function $M/H^{1/\delta }\equiv m(x)$ with respect to
$x$ alone can not be determined directly by this fit, since neither argument
of $M(x,1)$ on the right-hand side of Eq.\ (\ref{scal_form}) is small in the
critical region. Following the procedure discussed in Ref.\ \cite{Sava} we
determine the {\em characteristic scaling function} by plotting the reduced
data $m(x)/m(0)$ vs. $x/x_{0}$, where $x_{0}$ are the values of the argument
where deviations from power-law behavior start to occur (related to the
amplitude ${\cal {B}}$ in the magnetization vs. temperature curve (see Fig.\
1b). The resulting plot of the data is shown in Fig.\ 3b. The scaling plot
confirms both the consistency of the measured critical exponents within the
quoted error bars, and determines the scaling function of the phase
transition.

\begin{figure}[h]
\epsfxsize=76mm\epsffile[24 55 478 548]{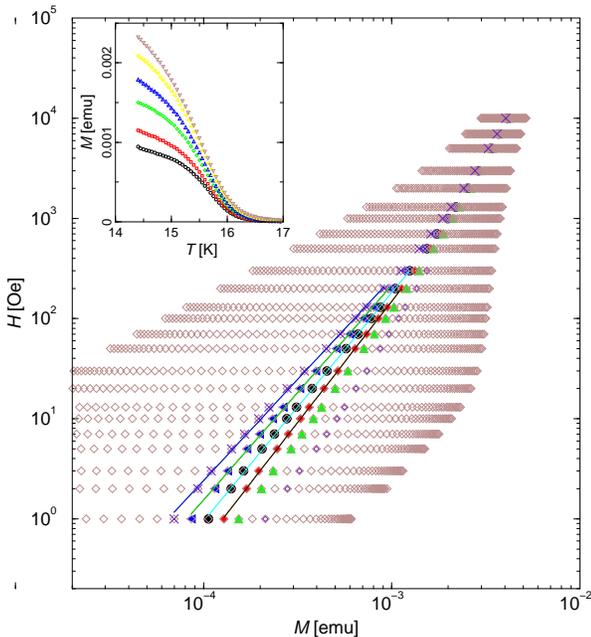}
\caption{Field dependence of magnetization for different temperatures in the
range from 14.4 K to 17.6 K (plotted right to left) every 0.05 K. Emphasized
are several isotherms near $T_c$, in particular for $T$ =16.1 K, 16.05 K, 16
K, 15.95 K and 15.9 K with the corresponding fit lines for low field values
also shown, their slopes giving the critical exponent $\protect\delta $. The
axes are cut off to reflect the uncertainty in the measurements. Inset:
Magnetization vs. temperature data for field values $H=$ 2, 3, 5, 7, 10, and
13 Oe.}
\end{figure}
Now let us comment on the obtained values for the critical exponents at the
ferromagnetic transition in TDAE-C$_{60}$ (a complete list of exponents is
given in Table\ 1). From previous measurements \cite{Blinc} of single
crystals it appears that TDAE-C$_{60}$ for 10K$\lesssim T<$16K is a
ferromagnet with localized magnetic moments with a very low anisotropy.
Therefore, we first discuss our results in comparison with the isotropic
three-dimensional Heisenberg model. The theoretical values calculated with
the Renormalization Group (RG) techniques \cite{RG} for Heisenberg model
with spins $S=1/2$ and nearest-neighbor interactions are also shown in
Table\ 1. It is clear that the measured exponents differ significantly from
the ones of $3d$ Heisenberg model. In addition, violation of the scaling
relation, i.e. $\gamma \neq \beta (\delta -1),$ indicates an entirely
different nature of the transition in TDAE-C$_{60}$. In particular, an
additional exponent $\bar{\gamma}\equiv \beta (\delta -1)$ can be defined,
which in turn violates the hyperscaling relation \cite{HSV} by an amount $%
\Omega $. A modified hyperscaling relation then holds:
\begin{equation}
2\beta +\gamma =(d+\Omega )\nu \ ,  \label{MHS}
\end{equation}
where $d=3$ is spatial dimension of the system and $\nu $ is the correlation
length exponent. Physical insight of the relation (\ref{MHS}) can be
achieved by considering another exponent $\bar{\beta}$ defined by \cite{DHSV}
\begin{equation}
{\bar{\beta}}/\nu =\beta /\nu -\Omega \ ,  \label{BB}
\end{equation}
such that, together with $\bar{\gamma}/\nu =\gamma /\nu +\Omega $ the
original HS relation is satisfied, i.e., $2\bar{\beta}+\bar{\gamma}=d\nu $.
>From the known $\bar{\gamma}$=0.96 we find$\ \Omega \nu =-0.26$ and thus $%
\bar{\beta}=1.01$. This immediately gives $\nu =(2\bar{\beta}+\bar{\gamma}%
)/3=0.99$. Therefore, $\Omega =-0.26<0$, meaning that the effective
dimension $d^{\prime }\equiv d+\Omega $ in Eq.\ (\ref{MHS}) is reduced by $%
\sim 1/4,$ within the error bars of the measurements. A reduced effective
dimension indicates that the fluctuations at the transition are stronger
than purely thermal fluctuations. For instance an enhancement of
fluctuations due to configurational disorder in random field systems such as
considered in ref. \cite{RFIM}. For instance, within the random-field Ising
model, the dimensional reduction for $d=3$ was estimated to lie between $%
\Omega =-1$ and $\Omega =-1.5$ \cite{HSV}. The origin of disorder in TDAE-C$%
_{60}$ can indeed be related to random spatial realizations of the contact
configurations of C$_{60}$ molecular orientations. As discussed in Ref.\
\cite{nature}, only the configuration with alternating $I$--$II$
orientations along $c$-axis is compatible with ferromagnetism, but other
contact configurations may also occur with a finite probability. Therefore,
it is quite plausible that the long-range ferromagnetic order sets-in when a
percolating cluster of the ``right'' contact configurations is established.
Indeed, the value of the exponent $\bar{\beta}\approx 1$ is quite compatible
with the backbone percolation \cite{percolation} of a random incipient
cluster.

Beside the measured exponents $\beta $, $\gamma $ and $\delta $ in Table 1
we have computed the rest of the exponents using the valid scaling
relations. Of course direct measurements of the exponents $\nu $, $\gamma $,
$\eta $, e.g., by scattering experiments, and measurements of the specific
heat exponent $\alpha $ in TDAE-C$_{60}$ are necessary in order to
independently confirm\ the accuracy of these values. Preliminary
measurements in fact suggest that no anomaly occurs in the specific heat\cite
{spec-heat}, in agreement with the predicted $\alpha <0$. Also, for
comparison, in Table\ 1 we have quoted the accepted values of the exponents
for random-field Ising systems as well as for the case of a random-exchange
(or diluted) Ising model\cite{REIM1}, in which disorder of the kind expected
here is a relevant perturbation at the phase transition. At this point we
wish to emphasize the distinction between the microscopic origin of the FM
interaction between C$_{60}$ molecules and the fluctuation mechanisms
leading to collective ferromagnetic behavior. The global response studied
here does not allow us to make detailed conclusions about the microscopic
picture, for which local microscopic probes such as ESR \cite{Mizoguchi}\ and
X-ray structure\cite{nature} are more appropriate.

\begin{figure}[h]
\epsfxsize=80mm\epsffile[26 72 506 336]{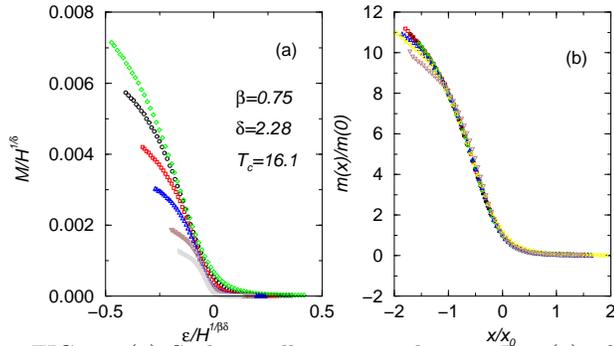}
\caption{(a) Scaling collapse according to Eq.\ (1) of the magnetization vs.
temperature curves for several values of the field $H=$ 2, 3, 5, 7, 10, and
13 Oe, which are shown in the inset to Fig.\ 2. $\protect\epsilon \equiv
(T-T_c)/T_c$ with $T_c=$ 16.1 and values of the exponents $\protect\beta %
=0.75$ and $\protect\delta = 2.28$ are used in the fit. The deviations from
the master curve behavior for different $H$ values indicate a crossover
value $x_0$ of $x\equiv \protect\epsilon/H^{1/\protect\beta \protect\delta }$
where the two variables in Eq.\ (1) become comparable. (b) The
characteristic scaling function $m(x)/m(0)$ vs. $x/x_0$ computed from data
shown in Fig.\ 3a as a function of the variable $x$.}
\end{figure}

\narrowtext
\begin{table}[th]
\caption{$^{a}$ Numerical values of the critical exponents of TDAE-C$_{60}$:
measured in this work $\protect\gamma $, $\protect\beta $, and $\protect%
\delta $, and the remaining exponents are computed via valid scaling
relations, as explained in the text. $\Omega $ represents effective
dimensional reduction in the hyperscaling relation. $^{b}$ Exponents of
random percolation, Ref.\ [17]. Here $\Omega $ and ${\bar{\protect\beta}}$
are the exponents related to the backbone percolation. $^{c}$ Critical
exponents for the random-field Ising model from first reference in [14]. $%
^{d}$ Exponents for the random-exchange Ising model from Ref.\ [20], and $%
^{e}$ for pure Heisenberg model, from Ref. \ [13], all for spatial dimension
$d=3$. }
\label{Table1}
\begin{center}
\begin{tabular}{|c||c|c|c|c|c||c|c|}
\hline
$Sys.$ & $\gamma$ & $\beta $ & $\delta $ & $\nu $ & $\alpha $ & $\Omega $ & $%
{\bar{\beta }}$ \\ \hline\hline
TDAE$^a$ & 1.22 & 0.75 & 2.28 & 1.06 & -0.72 & -0.26 & 1.01 \\ \hline
RP$^b$ & 1.82 & 0.41 & 5.43 & 0.88 & - & -0.66 & 0.99 \\ \hline
RFIM$^c$ & 1.9 & 0.06 & ? & 1.02 & -0.02 & -1 & 1.06 \\ \hline
REIM$^d$ & 1.34 & 0.35 & 4.78 & 0.68 & -0.05 & 0 & - \\ \hline
PHM$^e$ & 1.38 & 0.36 & 4.80 & 0.70 & -0.12 & 0 & - \\ \hline
\end{tabular}
\end{center}

To conclude, the excellent reproducibility of the measured exponents in
different crystals and over time strongly suggests that the exponents are
intrinsic to the material. Violation of the hyperscaling relation that
implies an effective dimensional reduction by approximately a quarter ($%
d^{\prime }\approx d-1/4$) indicates that additional degrees of
freedom---rotation of the C$_{60}$ molecules---significantly alters the
nature of the ferromagnetic phase transition in TDAE-C$_{60}$. The
transition appears to be in a new universality class which shares some
similarity with a backbone percolation and is attributed to the presence of
disorder in the C$_{60}$ molecular orientations near the transition
temperature.

\acknowledgments
This work was supported by the Ministry of education, science and sport of
the Republic of Slovenia. The authors wish to acknowledge the ESF\ MOLMAG
program and the Japanese STA fellowship for supporting part of this work.

\end{table}
\end{multicols}


\begin{references}
\bibitem{comment_1}  The extrapolation is reliable for small deviations of
temperature from $T_{c}$, where various curves are linear in a
double-logarithmic plot.
\end{references}
\end{document}